\documentclass[twocolumn,preprintnumbers,prl]{revtex4-1}
\pdfoutput=1

\usepackage{amsmath,amssymb,bm,mathrsfs}
\usepackage{hyperref}
\usepackage{url}
\usepackage{breakurl}
\usepackage{graphicx}

\def\app#1#2{%
  \mathrel{%
    \setbox0=\hbox{$#1\sim$}%
    \setbox2=\hbox{%
      \rlap{\hbox{$#1\propto$}}%
      \lower1.1\ht0\box0%
    }%
    \raise0.25\ht2\box2%
  }%
}
\def\approxprop{\mathpalette\app\relax}

\begin{document}

\preprint{UCB-PTH-14/32}

\title{Supersymmetry from Typicality}

\author{Yasunori Nomura and Satoshi Shirai}
\affiliation{Berkeley Center for Theoretical Physics, Department of Physics, 
  University of California, Berkeley, CA 94720}
\affiliation{Theoretical Physics Group, Lawrence Berkeley National 
  Laboratory, Berkeley, CA 94720}

\begin{abstract}
We argue that under a set of simple assumptions the multiverse leads 
to low energy supersymmetry with the spectrum often called spread or 
mini-split supersymmetry:\ the gauginos are in the TeV region with 
the other superpartners two or three orders of magnitude heavier. 
We present a particularly simple realization of supersymmetric grand 
unified theory using this idea.
\end{abstract}

\maketitle

\section{Introduction}

Supersymmetry is an elegant extension of spacetime symmetry, which arises 
naturally in string theory---the leading candidate for the fundamental 
theory of quantum gravity.  A striking property of supersymmetry is 
its high capability to control quantum corrections; in particular, 
it can protect the mass of a scalar field, such as the Higgs field, 
which has sizable non-derivative interactions.  This property was 
used to argue that the supersymmetric partners of the standard model 
particles are expected at the weak scale:\ the well-known naturalness 
argument~\cite{Martin:1997ns}.  On the other hand, the Large Hadron 
Collider (LHC) experiment has not seen any sign of superpartners so far, 
which is beginning to threaten this argument~\cite{Hall:2011aa}.

The plethora of string theory vacua~\cite{Bousso:2000xa} suggests that 
the naive naturalness argument should be modified.  Because of the eternal 
nature of inflation~\cite{Guth:1982pn}, all these vacua are expected to 
be physically populated, leading to the picture of the multiverse (or 
quantum many-universes~\cite{Nomura:2011dt}).  In this picture, our 
universe is one of the many universes in which low-energy physical laws 
take different forms, and the naturalness argument is replaced with the 
{\it typicality} argument~\cite{Vilenkin:1994ua}:\ we are typical observers 
among all the observers in the multiverse.  Specifically, the probability 
that physical parameters $x_i$ ($i=1,2,\cdots$) are observed between 
$x_i$ and $x_i + dx_i$ is given by $P(\{x_i\})\, d\{x_i\}$ with
\begin{equation}
  P(\{x_i\}) \propto \int\! d\{y_j\}\, 
    f\bigl(\{x_i\},\{y_j\}\bigr)\, n\bigl(\{x_i\},\{y_j\}\bigr),
\label{eq:prob}
\end{equation}
where $y_j$ ($j=1,2,\cdots$) are parameters other than $x_i$ which vary 
independently of $x_i$'s; $f$ is the a priori distribution function 
determined by the statistics of the landscape of vacua and their population 
dynamics, while $n$ is the anthropic weighting factor representing 
the probability of finding observers for a given $\{ x_i, y_j \}$. 
As is well known, this potentially allows us to understand the smallness 
of the cosmological constant (or dark energy) observed in our 
universe~\cite{Weinberg:1987dv,Martel:1997vi}.

In this letter, we consider what Eq.~(\ref{eq:prob}) implies for 
the masses of superpartners.  (For earlier studies, see e.g.\ 
\cite{Banks:2003es,Douglas:2004qg}.)  We argue that under a set of 
simple assumptions the multiverse leads to the spectrum often called 
spread or mini-split supersymmetry, which has attracted renewed interest 
recently~\cite{Hall:2011jd,Ibe:2011aa,Arvanitaki:2012ps,Hall:2012zp,%
ArkaniHamed:2012gw}:\ the gauginos are in the TeV region with the other 
superpartners two or three orders of magnitude heavier.  (For earlier work, 
see~\cite{Giudice:1998xp,Wells:2003tf,ArkaniHamed:2006mb,Acharya:2008bk}.) 
We find it encouraging that the typicality (or refined naturalness) 
argument, suggested by the fundamental theory, can lead---under rather 
simple assumptions---to a superpartner spectrum that does not have 
tension with the LHC results so far while allowing the possibility for 
future discovery.  Note that this spectrum preserves the successful 
prediction of supersymmetric gauge coupling unification, and yet does 
not suffer from (or at least highly ameliorates) problems associated 
with low energy supersymmetry, such as the supersymmetric flavor and 
$CP$ problems and the cosmological gravitino problem.

Throughout the letter, we assume that the electroweak scale is anthropically 
selected in the multiverse.  While the physical effects that are 
responsible for this selection are not fully understood, we at least 
know that changing it by a factor of a few leads to drastic changes 
of the universe~\cite{Agrawal:1997gf}.  We therefore take
\begin{equation}
  n\bigl(\{x_i\},\{y_j\}\bigr) \approxprop \delta(v-v_{\rm obs}),
\label{eq:v-fix}
\end{equation}
in Eq.~(\ref{eq:prob}), and see what distribution of the superpartner 
mass scale $P(\tilde{m} \in \{x_i\})$ is obtained.  Here, $v$ is 
the vacuum expectation value (VEV) of the standard model Higgs field 
calculated in terms of parameters $\{ x_i, y_j \}$, and $v_{\rm obs} 
\simeq 174~{\rm GeV}$ is the observed value (in units of some mass scale 
that is fixed in the analysis).  In our analysis, we vary essentially 
only parameters directly relevant for our question:\ the overall 
supersymmetry breaking mass scale $\tilde{m}$, the supersymmetric 
Higgs mass $\mu$, and the VEV of the superpotential $W_0$, which is 
needed to cancel the cosmological constant.  We do not expect, however, 
that our basic conclusion is sensitive to this restriction.

A key element that characterizes our scenario is the assumption that 
the supersymmetric Higgs mass term (the $\mu$ term)---which is the only 
relevant supersymmetric operator in the $R$-parity conserving minimal 
supersymmetric standard model (MSSM)---is not protected by any artificial 
symmetry.  This implies that the theoretically natural size of $\mu$ is 
of $O(M_*)$, where $M_*$ is the cutoff scale of the MSSM, and $\mu$ takes 
a value close to $\tilde{m}$ only because of the anthropic selection of 
the weak scale.  This provides a simple environmental solution to the 
$\mu$ problem.  Together with the assumptions that supersymmetry is broken 
dynamically and that the supersymmetry breaking field is not a singlet, 
we find that the claimed spectrum is obtained after applying all the 
selection effects, especially that associated with the abundance of 
dark matter.  This basic argument will be presented in the next section.

A particularly interesting realization of our scenario is obtained 
in the context of the minimal supersymmetric grand unified theory 
(GUT)~\cite{Dimopoulos:1981zb}.  In this theory, the $\mu$ term much 
smaller than the unified scale is obtained only as a result of a fine 
cancellation between different contributions, so that the setup relevant 
for our scenario is realized automatically.  Despite the minimality 
of the model, there is no problem of doublet-triplet splitting or 
unacceptably fast dimension-5 proton decay.  The precision of gauge 
coupling unification is also improved compared with conventional weak 
scale supersymmetry.  This, therefore, provides one of the most attractive 
realizations of supersymmetric GUT.

At the end of this letter, we also comment on the case in which 
the supersymmetry breaking field is a singlet.  In this case, the 
superpartner masses can be at a high energy scale with the VEVs of 
the two Higgs doublets being almost equal.  This allows for realizing 
the high scale supersymmetry scenario discussed in Ref.~\cite{Hall:2009nd}.

\section{Basic Argument}

We postulate that in the landscape of possible theories in the multiverse, 
probabilities relevant for observers are dominated by the branch having 
the following properties:
\begin{itemize}
\item[(i)]
Low energy theory below some high scale $M_*$ is the MSSM with $R$-parity 
conservation (possibly with the QCD axion supermultiplet to solve the 
strong $CP$ problem).  We are agnostic about the precise nature of the 
scale $M_*$ here, except that we assume it is not many orders of magnitude 
smaller than the reduced Planck scale $M_{\rm Pl}$.  Later we will 
consider the case in which $M_*$ is identified as the unification 
scale of supersymmetric GUT.
\item[(ii)]
Supersymmetry is broken dynamically as a result of dimensional transmutation 
associated with some hidden sector gauge group so that $\tilde{m} \propto 
e^{-8\pi^2/c g^2}$~\cite{Witten:1981nf}, where $g$ and $c$ are the hidden 
sector gauge coupling and an $O(1)$ coefficient, respectively.
\item[(iii)]
The superfield $X$ responsible for supersymmetry breaking, $\langle 
X|_{\theta^2} \rangle = F_X$, is not a singlet.  This prohibits the terms 
$[X H_u H_d]_{\theta^2}$ and $[X {\cal W}^\alpha {\cal W}_\alpha]_{\theta^2}$ 
to appear in the Lagrangian at tree level, where $H_{u,d}$ and 
${\cal W}_\alpha$ represent the two Higgs doublet and gauge strength 
superfields of the MSSM, respectively.
\end{itemize}

In addition to these somewhat ``standard'' assumptions, we also postulate 
that the branch in the landscape dominating the probabilities has the 
property:
\begin{itemize}
\item[(iv)]
There is no ``artificial'' symmetry below $M_*$ controlling the size 
of operators in the MSSM (except for approximate, and perhaps accidental, 
flavor symmetries associated with the smallness of the Yukawa couplings). 
In particular, there is no approximate global symmetry that dictates 
the values of $\mu$ and $W_0$, such as the ``Peccei-Quinn'' symmetry 
$H_{u,d} \rightarrow e^{i\alpha} H_{u,d}$, $CP$ symmetry, or a continuous 
$R$ symmetry.
\end{itemize}
As we will see, this implies that the $\mu$ problem is solved by the 
anthropic selection associated with electroweak symmetry breaking. 
In the fundamental picture in the landscape, this corresponds to 
postulating that having a theoretical mechanism of suppressing the 
$\mu$ term ``costs'' more than the fine-tuning needed to make $|\mu|$ 
small to obtain acceptable electroweak symmetry breaking.

The postulate~(iii) above implies that at the leading order, supersymmetry 
breaking masses in the MSSM sector arise from operators of the form 
$[X^\dagger X \Phi^\dagger \Phi/\Lambda^2]_{\theta^4}$ and $[X^\dagger X 
H_u H_d/\Lambda^2]_{\theta^4}$, where $\Phi$ represents the MSSM matter 
and Higgs fields, as well as from tree-level supergravity effects. 
This yields
\begin{equation}
  m_{\tilde{f},H_u,H_d}^2 = c_{\tilde{f},H_u,H_d} \tilde{m}^2,
\qquad
  b = c_b \tilde{m}^2 - \mu m_{3/2},
\label{eq:soft-1}
\end{equation}
where $\tilde{m} \equiv |F_X/\Lambda|$, $m_{3/2} = F_X/\sqrt{3} M_{\rm Pl}$ 
is the gravitino mass, and $m_{\tilde{f},H_u,H_d}^2$ represent the 
(non-holomorphic) supersymmetry breaking squared masses for the sfermion 
and Higgs fields while $b$ is the holomorphic supersymmetry breaking 
Higgs mass squared.  $c_{\tilde{f},H_u,H_d,b}$ are coefficients, which 
we take to scan for values of order unity in the landscape, and $\Lambda$ 
is the mediation scale of supersymmetry breaking, which we take to be 
roughly of the order of---e.g.\ within an order of magnitude of---the 
reduced Planck scale:\ $\Lambda \sim M_{\rm Pl}$.

We now discuss the probability distribution of $\tilde{m}$, $\mu$, and 
$W_0$ in this setup.  With the assumption in (ii), it is reasonable to 
expect that the a priori distribution of $\tilde{m}$ is approximately 
flat in logarithm:\ $f\, d\tilde{m} \approxprop d\ln \tilde{m}$.  On 
the other hand, the assumption in (iv) implies that the distribution of 
$\mu$ is given by $f\, d\mu \propto d\mathrm{Re}\,\mu\, d\mathrm{Im}\,\mu 
\propto |\mu| d|\mu| d{\rm arg}(\mu)$ for $|\mu| \lesssim M_*$, and 
similarly for $W_0$.  We thus take $\{ x_i \} = \tilde{m}$ and $\{ y_j \} 
= \{ |\mu|, |W_0|, {\rm arg}(\mu), {\rm arg}(W_0), c_{\tilde{f},H_u,H_d}, 
c_b \}$ with
\begin{equation}
  f \propto \frac{1}{\tilde{m}}\, |\mu|\, |W_0|,
\label{eq:apriori}
\end{equation}
and study what values of these parameters are selected by anthropic 
conditions.  In our analysis, we ignore possible variations of all the 
other parameters of the theory, but we do not expect that our basic 
conclusion is overturned when the full variations are performed with 
appropriate anthropic conditions.  This approach is analogous to that 
adopted in the original argument for the cosmological constant in 
Ref.~\cite{Weinberg:1987dv}.

Let us first isolate the selection effects from the weak scale and the 
cosmological constant by writing
\begin{equation}
  n \approx \delta(v-v_{\rm obs})\, 
    \theta(\rho_{\Lambda,{\rm max}} - |\rho_\Lambda|)\, 
    \hat{n}(\tilde{m},\mu),
\label{eq:n-isolate}
\end{equation}
where $v$ and $\rho_\Lambda$ are the values of the Higgs VEV and the 
vacuum energy density calculated in terms of $\tilde{m}$, $\mu$, and 
$W_0$; $\rho_{\Lambda,{\rm max}} = |\gamma \rho_{\Lambda,{\rm obs}}|$ 
is the anthropic upper bound on the value of $\rho_\Lambda$, which 
we have taken symmetric around $\rho_\Lambda = 0$ for simplicity, and 
$\gamma$ is a constant not too far from order unity.  We expect that 
the residual anthropic weighting factor, $\hat{n}$, does not depend 
strongly on the values of $W_0$ or $c_{\tilde{f},H_u,H_d,b}$ in the 
relevant parameter region, which we assume to be the case.

By integrating over $W_0$ in Eq.~(\ref{eq:prob}), using the expression 
$\rho_\Lambda = |F_X|^2 - 3|W_0|^2/M_{\rm Pl}^2$ with $|F_X| = \tilde{m} 
\Lambda$, we obtain
\begin{equation}
  P(\tilde{m}) \approxprop \int\! d|\mu|\, d{\rm arg}(\mu)\, d\{ c_i \}\,
    \frac{1}{\tilde{m}}\, |\mu|\, \delta(v-v_{\rm obs})\, 
    \hat{n}(\tilde{m},\mu),
\label{eq:after-cc}
\end{equation}
where $\{ c_i \} \equiv c_{\tilde{f},H_u,H_d,b}$, and $v$ depends 
on $\tilde{m}$, $\mu$, and $\{ c_i \}$.  An important point 
is that the integration of $W_0$ does not provide an extra 
probability bias for $\tilde{m}$, $|\mu|$, ${\rm arg}(\mu)$, or 
$\{ c_i \}$~\cite{Douglas:2004qg}---the effective a priori distribution 
function for these parameters is still given by $f_{\rm eff} \propto 
|\mu|/\tilde{m}$.  The value of $W_0$ is selected to be $|W_0| = |F_X| 
M_{\rm Pl}/\sqrt{3}$ with the phase, ${\rm arg}(W_0)$, unconstrained. 
This leads to the gravitino mass roughly comparable to the sfermion 
masses:
\begin{equation}
  m_{3/2} \sim \tilde{m}.
\label{eq:m32}
\end{equation}
In fact, we may naturally expect that $\Lambda \lesssim M_{\rm Pl}$, 
so that $m_{3/2}$ is comparable to or somewhat (e.g.\ up to an order of 
magnitude) smaller than $\tilde{m}$.

The selection from electroweak symmetry breaking acts on the mass-squared 
matrix of the doublet Higgs bosons
\begin{equation}
  {\cal M}_H^2 = \left( \begin{matrix}
    |\mu|^2 + m_{H_u}^2 & b \\
    b^* & |\mu|^2 + m_{H_d}^2
  \end{matrix} \right),
\label{eq:M_H-sq}
\end{equation}
where $m_{H_{u,d}}^2 \approx O(\tilde{m}^2)$ and $|b| \approx O({\rm max}\{ 
\tilde{m}^2, |\mu| \tilde{m} \})$.  It requires the smallest eigenvalue 
of ${\cal M}_H^2$ to be $\approx -v_{\rm obs}^2$.  For $\tilde{m} \ll 
v_{\rm obs}$, it is not possible to satisfy this requirement.  For 
$\tilde{m} \gtrsim v_{\rm obs}$, on the other hand, the requirement 
can be met for some values of $\{ c_i \}$ if $|\mu| \lesssim \tilde{m}$, 
while a value of $|\mu|$ much larger than $\tilde{m}$ makes the 
eigenvalues of ${\cal M}_H^2$ both positive regardless of $\{ c_i \}$. 
This therefore selects the value of $|\mu|$ to be
\begin{equation}
  |\mu| \approx \tilde{m},
\label{eq:mu}
\end{equation}
since the probability of having $|\mu| \ll \tilde{m}$ is suppressed by 
the fact that $f_{\rm eff} \approxprop |\mu|$.  More explicitly, for 
a fixed $\tilde{m}$, the probability distribution for $|\mu|$ is given 
roughly by $P(|\mu|; \tilde{m})\, d|\mu| \approxprop |\mu| \theta(\tilde{m} 
- |\mu|)\, d|\mu|$.  We thus find that by integrating over $|\mu|$, 
${\rm arg}(\mu)$, and $\{ c_i \}$ in Eq.~(\ref{eq:after-cc}), we obtain
\begin{equation}
  P(\tilde{m})\, d\tilde{m} \approxprop 
    \left\{ \begin{array}{ll} 
      0\; & \mbox{for } \tilde{m} \lesssim v_{\rm obs}, \\
      n_{\rm eff}(\tilde{m})\, d\ln \tilde{m}\; 
        & \mbox{for } \tilde{m} \gtrsim v_{\rm obs},
    \end{array} \right.
\label{eq:P-tildem}
\end{equation}
where $n_{\rm eff}(\tilde{m}) \equiv \hat{n}(\tilde{m},\mu = \tilde{m})$, 
and we have assumed that the anthropic weighting factor $\hat{n}$ 
does not depend strongly on the precise value of $\mu$.  Interestingly, 
for $\tilde{m} \gtrsim v_{\rm obs}$, the preference to smaller 
values of $\tilde{m}$ by electroweak fine-tuning, $P \propto 
v_{\rm obs}^2/\tilde{m}^2$, is exactly canceled by the a priori 
distribution of $\mu$, which through Eq.~(\ref{eq:mu}) prefers larger 
values of $\tilde{m}$, i.e.\ $P \propto |\mu| d|\mu| \sim \tilde{m}^2 
d\ln\tilde{m}$.  There is, therefore, no net preference for the 
scale of superpartner masses before a further anthropic selection, 
$n_{\rm eff}(\tilde{m})$, is applied.

What function should we consider for $n_{\rm eff}(\tilde{m})$?  Note that 
with Eqs.~(\ref{eq:m32}) and (\ref{eq:mu}), the pattern of the spectrum 
of the superpartners and the MSSM Higgs bosons ($H^0$, $H^\pm$, and $A^0$) 
is fixed in terms of the single mass parameter $\tilde{m}$:\ in order 
of decreasing masses
\begin{equation}
  M_{\tilde{f}} \approx M_{\tilde{h}} \approx \tilde{m},
\qquad
  M_{H^{0,\pm},A^0} \approx \tilde{m},
\label{eq:spectrum-1}
\end{equation}
where $\tilde{h}$ represents the Higgsinos,
\begin{equation}
  M_{\tilde{G}} = \epsilon \tilde{m};
\quad
  \epsilon \approx O(0.1~\mbox{--}~1),
\label{eq:spectrum-2}
\end{equation}
where $\tilde{G}$ is the gravitino, and the lightest set of superpartners 
are the gauginos with masses
\begin{equation}
  M_a = \frac{b_a g_a^2}{16\pi^2} m_{3/2} + \frac{d_a g_a^2}{16\pi^2} L,
\label{eq:spectrum-3}
\end{equation}
($a=1,2,3$), which are generated by anomaly mediation~\cite{Randall:1998uk,%
Giudice:1998xp} (the first term) and Higgsino-Higgs loops (the second 
term).  Here, $g_a$ are the standard model gauge couplings (in the SU(5) 
normalization for U(1) hypercharge), $(b_1, b_2, b_3) = (33/5, 1, -3)$, 
$(d_1, d_2, d_3) = (3/5, 1, 0)$, and
\begin{equation}
  L \approx O(\tilde{m} \sin(2\beta)),
\label{eq:L}
\end{equation}
where $\tan\beta \equiv \langle H_u \rangle/\langle H_d \rangle$. 
Our question is:\ what environmental effects do we expect when we vary 
$\tilde{m}$ keeping the relations in Eqs.~(\ref{eq:m32}) and (\ref{eq:mu})?

In general, there could be many subtle effects associated with a variation 
of $\tilde{m}$.  For example, for fixed high energy gauge and Yukawa 
couplings, varying $\tilde{m}$ leads to changes of low energy parameters 
such as the QCD scale and the quark and lepton masses.  These changes, 
however, may be compensated by the corresponding variations of the high 
energy parameters, and their full analysis will require detailed knowledge 
of the statistical properties of the landscape beyond the scope of this 
letter.  Below, we will focus on what is arguably the dominant environmental 
effect of varying $\tilde{m}$:\ the change of the relic abundance of 
the lightest supersymmetric particle (LSP) contributing to dark matter 
of the universe.

\begin{figure}[t]
  \includegraphics[clip, width=0.45 \textwidth]{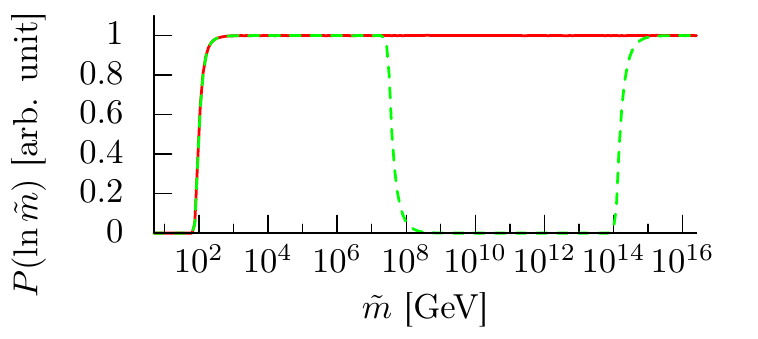}
\caption{A schematic depiction for the probability distribution 
 of $\tilde{m}$.  The red solid (green dashed) line represents 
 the distribution before (after) the dark matter condition, 
 $n_{\rm eff}(\tilde{m})$, is imposed.  Here, we have assumed 
 $T_{\rm R} = 10^9~{\rm GeV}$ and that the anthropic upper bound 
 of the dark matter abundance is at $\Omega_{\rm DM} h^2 = 100$.}
\label{fig:plot}
\end{figure}

As discussed in Ref.~\cite{Hall:2011jd}, this effect leads to a large 
``forbidden window'' for the value of $\tilde{m}$, in which the LSP 
relic abundance far exceeds the observed dark matter abundance and 
$n_{\rm eff}(\tilde{m}) \approx 0$.  While the precise nature of the 
anthropic upper bound on the dark matter abundance is not well understood, 
we expect that there is some upper bound.  This upper bound may not be 
sharp or close to the observed value (for discussions on possible upper 
bounds, see Ref.~\cite{Tegmark:2005dy}), but it still excludes a large 
region of $\tilde{m}$ between some values $\tilde{m}_1$ and $\tilde{m}_2$. 
For the spectrum in Eqs.~(\ref{eq:spectrum-1}~--~\ref{eq:spectrum-3}), 
the LSP is the wino in most of the parameter space, whose relic abundance 
has both thermal and non-thermal contributions; see Ref.~\cite{Hall:2012zp} 
for a detailed analysis.  (If $\epsilon$ in Eq.~(\ref{eq:spectrum-2}) 
is small, the second term in Eq.~(\ref{eq:spectrum-3}) may dominate and 
the bino can be the LSP; the allowed dark matter window in this case is 
small~\cite{Hall:2012zp}.)  The probability distribution for the superpartner 
mass scale $\tilde{m}$ is then given by Eq.~(\ref{eq:P-tildem}) as
\begin{equation}
  P(\tilde{m})\, d\tilde{m} \approx 
    \left\{ \begin{array}{ll} 
      {\cal C}\, d\ln \tilde{m}\; 
        & \mbox{for } v_{\rm obs} \lesssim \tilde{m} \lesssim \tilde{m}_1,\; 
          \tilde{m} \gtrsim \tilde{m}_2, \\
      0\; & \mbox{for other values of } \tilde{m},
    \end{array} \right.
\label{eq:P-tildem-2}
\end{equation}
where ${\cal C}$ is the normalization constant.  A reasonable (though 
highly uncertain) estimate for $\tilde{m}_1$ is given by
\begin{equation}
  \tilde{m}_1 \approx O(10^6~\mbox{--}~10^8~{\rm GeV}),
\label{eq:tildem_1}
\end{equation}
and $\tilde{m}_2$ is given by the value of $\tilde{m}$ in which the 
LSP becomes sufficiently heavier than the reheating temperature after 
inflation $T_{\rm R}$, which we assume to take some high scale value, 
e.g.\ $T_{\rm R} \gtrsim 10^9~{\rm GeV}$ (possibly selected by the 
requirement of baryogenesis).  The resulting distribution is depicted 
schematically in Fig.~\ref{fig:plot}.

With Eq.~(\ref{eq:P-tildem-2}), it is easy to imagine that $\tilde{m}$ 
is selected randomly in the logarithmic scale in the region between 
$v_{\rm obs} \approx O(100~{\rm GeV})$ and $\tilde{m}_1$.  In 
particular, this can lead to the spread or mini-split type spectrum 
in Eqs.~(\ref{eq:spectrum-1}~--~\ref{eq:spectrum-3}) with 
\begin{equation}
  \tilde{m} \sim m_{3/2} \approx O(10~\mbox{--}~1000~{\rm TeV}),
\label{eq:fin-scale}
\end{equation}
as a typical spectrum one observes in the multiverse.  The phenomenology 
of this spectrum is discussed in detail in Ref.~\cite{Hall:2012zp} and 
references therein.  In particular, the observed Higgs boson mass of 
$m_{h^0} \simeq 126~{\rm GeV}$ can be easily accommodated, and dark matter 
can be composed of a mixture of the wino and the QCD axion (and possibly 
other components as well) with some unknown ratio, which depends on the 
statistical distribution of the axion decay constant, among other things.

\section{Minimal GUT Realization}

A particularly attractive and solid realization of our scenario 
is obtained in the context of the minimal supersymmetric SU(5) 
GUT~\cite{Dimopoulos:1981zb}.  In this theory, the SU(5) symmetry 
is broken by the VEV of $\Sigma(24)$.  The superpotential relevant 
for the GUT breaking is
\begin{equation}
  W_\Sigma = \frac{m_\Sigma}{2} \mathrm{tr} \Sigma^2 
    + \frac{\lambda_\Sigma}{3} \mathrm{tr} \Sigma^3,
\label{eq:W_Sigma}
\end{equation}
leading to $\langle \Sigma \rangle = (m_\Sigma/\lambda_\Sigma) 
\mathrm{diag}(2,2,2,-3,-3)$.  For a non-singlet supersymmetry breaking 
field $X$, the shift of this vacuum due to supersymmetry breaking effects 
is small.  The superpotential for the Higgs fields, $H_5 = (H_C, H_u)$ 
and $\bar{H}_5 = (\bar{H}_C, H_d)$, is given by
\begin{equation}
  W_H = \bar{H}_5 ( m_H + \lambda_H \Sigma ) H_5.
\label{eq:W_H}
\end{equation}
With the $\Sigma$ VEV, this leads to the supersymmetric masses for the 
MSSM Higgs doublets $H_{u,d}$ and their GUT partners $H_C, \bar{H}_C$:
\begin{equation}
  \mu = m_H - 3 \frac{\lambda_H m_\Sigma}{\lambda_\Sigma},
\qquad
  \mu_C = m_H + 2 \frac{\lambda_H m_\Sigma}{\lambda_\Sigma}.
\label{eq:mu-GUT}
\end{equation}
Since both terms in the right-hand side are of the order of the unification 
scale $M_{\rm unif} \approx O(10^{16}~{\rm GeV})$, having $|\mu| \ll 
M_{\rm unif}$ requires a large fine-tuning of parameters.  With this 
fine-tuning, $\mu_C$ is of order $M_{\rm unif}$.

The situation described above is exactly the one needed to realize our 
scenario.  In fact, we may apply the argument in the previous section 
without modification by identifying $M_*$ as the unification scale 
$M_{\rm unif}$.  This realization has several virtues from the GUT 
point of view as well.  First, the notorious doublet-triplet splitting 
problem is automatically ``solved'' because of the environmental selection 
for electroweak symmetry breaking.  Second, the precision of gauge 
coupling unification is also improved with the superpartner spectrum 
discussed here~\cite{Hisano:2013cqa}, compared with the case in conventional 
weak scale supersymmetry.  The dangerous dimension-5 proton decay caused 
by exchange of $H_C, \bar{H}_C$ is also suppressed because of the large 
sfermion masses, if the $1$-$3$ element of the doublet squark mass-squared 
matrix is sufficiently small~\cite{Hisano:2013exa}.  (This is ensured if 
the low energy theory has an appropriate flavor structure analogous to 
that in the Yukawa couplings, which also suppresses possible contributions 
from cutoff-scale suppressed tree-level operators~\cite{Dine:2013nga}.) 
This scenario, therefore, provides one of the most simple and attractive 
realizations of supersymmetric GUT.

\section{Comment on Singlet {\boldmath $X$}}

We finally comment on the case in which the assumption~(iii) is modified:\ 
the supersymmetry breaking field $X$ is a singlet.  In this case, the 
direct coupling of $X$ to the Higgs fields $W \supset -\lambda X H_u H_d$ 
is allowed, yielding $|b| = |\lambda| \Lambda \tilde{m}$.  By integrating 
over $\lambda \in \{ y_j \}$ in the expression in Eq.~(\ref{eq:prob}) 
using the electroweak symmetry breaking condition $\delta(v-v_{\rm obs})$, 
we find that the probability in the $\ln\tilde{m}$-$\ln|\mu|$ plane is 
peaked at $|\mu|^2/\Lambda \tilde{m} \approx |\lambda|_{\rm max}$; more 
explicitly,
\begin{equation}
  P(\ln\tilde{m}, \ln|\mu|) 
  \approxprop \frac{|\mu|^4}{(\Lambda \tilde{m})^2}\, 
    \theta\biggl( |\lambda|_{\rm max} - \frac{|\mu|^2}{\Lambda \tilde{m}} 
    \biggr)\, \hat{n}(\tilde{m},|\mu|),
\label{eq:singlet-2}
\end{equation}
where $|\lambda|_{\rm max}$ is the largest value of $|\lambda|$ 
in the landscape, which we expect is not too far from $O(1)$, and 
$\hat{n}(\tilde{m},|\mu|)$ is the anthropic weighting factor without 
including that for electroweak symmetry breaking.  

An important element in $\hat{n}(\tilde{m},|\mu|)$ arises from the 
fact that for $|b| \simeq |\mu|^2 \gg \tilde{m}^2$, a Higgs loop makes 
the mass squared for the top squark negative $m_{\tilde{t}}^2 \sim 
-(y_t^2/16\pi^2) |\mu|^2$.  Assuming that such a region, $\tilde{m}^2 
\lesssim 10^{-(2-3)} |\mu|^2$, is environmentally disallowed, we find
\begin{equation}
  \tilde{m} \gtrsim 10^{-(2-3)} |\lambda|_{\rm max} \Lambda,
\label{eq:singlet-3}
\end{equation}
so we expect that the superpartner mass scale is high.  Moreover, we 
find that
\begin{equation}
  \tan\beta - 1 \approx O\biggl(\frac{\tilde{m}^2}{|\mu|^2}\biggr) 
  \approx O\biggl(\frac{\tilde{m}}{|\lambda|_{\rm max} \Lambda}\biggr) 
  \ll 1,
\end{equation}
in most of the selected parameter space.  This, therefore, provides 
a simple realization of the high scale supersymmetry scenario discussed 
in Ref.~\cite{Hall:2009nd} with $m_{h^0} \simeq 126~{\rm GeV}$.  A more 
detailed analysis of the singlet $X$ case will be presented elsewhere.

\section*{Acknowledgments}

This work was supported in part by the Director, Office of Science, Office 
of High Energy and Nuclear Physics, of the U.S.\ Department of Energy under 
Contract DE-AC02-05CH11231 and in part by the National Science Foundation 
under grants PHY-0855653 and PHY-1214644.

\end{document}